\documentclass[conference]{IEEEtran}
\IEEEoverridecommandlockouts

\usepackage{cite}
\usepackage{amsmath,amssymb,amsfonts}
\usepackage{algorithmic}
\usepackage{graphicx}
\usepackage{textcomp}
\usepackage{subfig}
\usepackage{url}
\def\BibTeX{{\rm B\kern-.05em{\sc i\kern-.025em b}\kern-.08em
    T\kern-.1667em\lower.7ex\hbox{E}\kern-.125emX}}

\begin{document}

\title{Sensitivity Analysis of Dynamic Line Rating for ACSR Conductors using IEEE-738
}

\author{
\IEEEauthorblockN{Shashank Singh, Ashish Kumar Mishra, Vinod M. P.}
\IEEEauthorblockA{
\textit{Grid Software}\\
\textit{Siemens Technology and Services Private Limited}\\
Bangalore, India\\
\{shashanksingh, mishra.ashish-kumar, vinod.mp\}@siemens.com
}
\and
\IEEEauthorblockN{Christian Romeis}
\IEEEauthorblockA{
\textit{Grid Software}\\
\textit{Siemens AG}\\
Nuremberg, Germany\\
christian.romeis@siemens.com
}
}

\maketitle

\renewcommand{\thefootnote}{}
\footnotetext{\copyright 2026 IEEE. Personal use of this material is permitted.  Permission from IEEE must be obtained for all other uses, in any current or future media, including reprinting/republishing this material for advertising or promotional purposes, creating new collective works, for resale or redistribution to servers or lists, or reuse of any copyrighted component of this work in other works.}
\renewcommand{\thefootnote}{\arabic{footnote}}

\begin{abstract}
Dynamic Line Rating (DLR) is a novel technique that enhances the utilization of transmission line capacity. It is nevertheless unclear how much measurement uncertainty in important environmental parameters affects the DLR calculation. Using the IEEE-738 standard, this paper presents a systematic parametric sensitivity analysis of DLR for a 795~kcmil ACSR Drake conductor. The DLR computation encompasses 832 operating points which includes daytime/nighttime solar scenarios, clear/non-clear atmospheric clearness conditions, wind speeds ranging from 0 to 15.25 m/s, and ambient temperatures between 15 and 50 $^\circ$C. According to sensitivity analysis, wind sensitivity tends to decrease at higher ambient temperatures, whereas temperature sensitivity tends to increase with wind speed. Pearson correlation analysis indicates a strong negative linear association between DLR and ambient temperature, and a strong positive linear association with wind speed. A regression model incorporates both individual and interactive effects of wind and temperature, and explains over 93\% of the observed DLR variability across all cases. Finally, the observations serve as a guide for operational planning and uncertainty assessment in DLR-based transmission systems.
\end{abstract}


\section{Introduction}
Overhead transmission lines are often described as electrical highways, enabling the transfer of large amounts of electrical power over long distances across interconnected grids. Electricity demand has been growing steadily in recent years because of the rapid expansion of digital infrastructure, data centers, electric vehicles, and AI-powered applications. Thus, the existing transmission networks are under more stress. Hence, efficient utilization of available transmission capacity has become an increasingly important concern for system operators.

Aluminum conductor steel-reinforced (ACSR) conductors are the backbone of our electrical transmission systems. Transmission companies prefer these conductors because they offer a favorable combination of electrical conductivity, mechanical strength, and economic cost. Traditionally, transmission lines are operated using static line ratings. Static ratings are conservative current limits provided by manufacturers or utilities, typically calculated assuming worst-case ambient conditions—high temperature and low wind speed. While this approach ensures safe operation, it often leads to underutilization of the transmission network, since actual environmental conditions are frequently less severe than the assumed worst case \cite{abas2024-2}.

Dynamic Line Rating (DLR) is a methodology for calculating transmission capacity based on the interaction of heating and cooling thermodynamics with conductor physical attributes and geography. Several industrial standards and guidelines describe DLR calculation methods and constraints. Among these, IEEE-738 \cite{ieee738_2023} and CIGRE Technical Brochure 601 \cite{cigre601_2014} are widely recognized industrial standards. DLR guides power system network or energy management system operators toward better utilization of electrical transmission infrastructure, thereby reducing congestion, deferring costly network upgrades, and enhancing overall system reliability.

Even though the DLR standards mentioned earlier provide a thorough framework for calculations, the DLR's sensitivity to environmental factors such as wind direction, speed, ambient temperature, and atmospheric clarity can vary significantly across operating environments. These parameters do not change consistently in real-world grid operation. For instance, while ambient temperature usually fluctuates more slowly, wind conditions are highly dynamic and can change quickly. Similar differences exist between transmission line segments in terms of solar heat gain and atmospheric clearness (clear vs. non-clear-sky conditions). These changes in environmental parameters cause non-uniform thermal constraints along the same transmission line. As a result, identical conductors may show varying DLR values at different points, with some line segments becoming more restrictive than others. Therefore, measuring the impact of changes in these critical environmental factors on DLR is very important for planning and operational purposes and serves as the foundation for the systematic sensitivity analysis in this paper.

According to \cite{abas2024-2}, which looks at the effect of extreme thermal conditions on DLR, high ambient temperatures can lower line capacity below static limits, with wind speed being the most important factor. In \cite{abdelkader2025}, the economic benefits of DLR under heatwave conditions are illustrated. 
The DLR implementation reduces the risk of cascading overload and significantly reduces operating costs when compared to static ratings. In \cite{dhanesh2025}, advanced forecasting techniques for DLR are described, in which temporal convolutional networks predict day-ahead ratings that exceed static limits by over 120\% during critical operating periods.

Experimental validation of DLR thermal models is reported in \cite{kanalik2019} using CIGRE TB 601. Laboratory and field measurements confirm the accuracy of the calculations on operating lines in Slovakia. The reliability of forecast-based DLR is addressed in \cite{markovits2025} through statistical correction of weather-parameter errors and the introduction of safety factors for conservative operation. A geospatial assessment framework in \cite{maswedza2025} shows that transmission lines with substantial DLR headroom can be identified using historical meteorological data, demonstrating DLR's suitability for alleviating network constraints. Laboratory measurement systems for the thermal characteristics of ACSR conductors are described in \cite{milovanovic2025}, providing experimental validation frameworks for DLR studies.

In order to improve dispatch accuracy in large-scale systems, \cite{pan2022} investigated integrating DLR into grid operation by combining fast DLR calculations with power transfer distribution factor (PTDF) corrections in real-time market-clearing models. In \cite{riba2025}, the author quantifies the impact of corona losses on DLR thermal models. Corona losses have a negligible effect on DLR calculations in typical circumstances but can result in substantial derating for contaminated conductors. According to \cite{su2025}, a hybrid renewable system capacity forecasting that incorporates DLR achieves better short-term prediction accuracy during high-wind periods. Additionally, \cite{abas2024} not only provides a comprehensive examination of DLR methodologies and implementation challenges but also positions DLR as a grid flexibility tool, and highlights areas that require further research.

The literature mentioned above has addressed many facets of DLR, such as forecasting, validation, and integration. Nevertheless, there is a lack of a systematic comparative characterization of how interactions between environmental parameters affect DLR under different atmospheric and solar conditions. In order to fill this gap, this paper uses the IEEE-738 standard to present a thorough multi-scenario analysis of DLR calculations for a 795~kcmil ACSR Drake conductor. The study looks at how wind speed and temperature affect DLR calculations under four different scenarios that are categorized by solar conditions (daytime/nighttime) and atmospheric clearness (clear/non-clear). The following are the main contributions of this paper:

\begin{enumerate}
\item Aggregated DLR variation analysis across environmental parameter ranges under solar and atmospheric clearness conditions, quantifying rating spreads caused by temperature and wind, and creating operational envelopes for each scenario.
\item Sensitivity analysis quantifying temperature and wind sensitivity coefficients under varying atmospheric and solar conditions, characterizing how DLR responds to one parameter, depending on the levels of other environmental factors.
\item Pearson correlation analysis determining the strength and nature of associations between DLR and environmental parameters across all atmospheric and solar scenarios.
\item Linear regression modeling with wind--temperature interaction terms fitted independently for each atmospheric clearness and solar condition, capturing individual, combined, and coupled environmental effects on DLR calculations.
\end{enumerate}

The remainder of this paper is organized as follows. Section~II describes the experimental setup, including the IEEE-738 thermal model, conductor parameters, and environmental conditions. Section~III presents the analysis methodology and results, covering aggregated DLR variation, sensitivity analysis, correlation analysis, and regression modeling. Section~IV concludes the paper with a summary of key findings.

\section{Experimental Setup}

This section describes the computational setup, conductor parameters, environmental assumptions, and simulation procedure used to analyze the effect of wind speed, ambient temperature, atmospheric clearness, and solar conditions (i.e., daytime, nighttime) on DLR. All DLR calculations are performed using the Siemens Gridscale X\textsuperscript{TM} DLR software \cite{siemens_dlr}. The resulting DLR dataset is then used to perform sensitivity, correlation, and regression analyses.

\subsection{Dynamic Line Rating Model}

The DLR software \cite{siemens_dlr} supports both industrial standards, \cite{ieee738_2023} and \cite{cigre601_2014}, for DLR calculation. This paper follows IEEE-738-2023 where steady-state heat balance equation \eqref{eq:heatbalance} is center to DLR calculation. 
\begin{equation}
q_c + q_r = q_s + q_j
\label{eq:heatbalance}
\end{equation}

Where $q_c$ is the convective heat loss, $q_r$ is the radiative heat loss, $q_s$ is the solar heat gain, and $q_j$ is the resistive heating. All heat losses and gains are expressed in watt-per-meter ($W/m$). Subsequently, the DLR is obtained by solving the equation \eqref{eq:heatbalance} and can be expressed using \eqref{eq:dlr}.
\begin{equation}
I_{\text{DLR}} = \sqrt{\frac{q_c + q_r - q_s}{R(T_{\max})}}
\label{eq:dlr}
\end{equation}

Here $I_{\text{DLR}}$ is the DLR in ampere ($A$) and $R(T_{\max})$ is the AC resistance of conductor at the maximum allowable conductor operating temperature expressed in $\Omega/m$.

\subsection{Conductor Parameters}

The conductor parameters considered in this study are conductor type, maximum allowable temperature, emissivity and absorptivity, geographic latitude and azimuth, altitude, and day of year. The parameters remain constant in all analyses. These parameters are taken from \cite{ieee738_2023} and summarized in Table~\ref{tab:conductor_params}.

\begin{table}[ht]
\caption{Conductor Parameters}
\label{tab:conductor_params}
\centering
\small
\begin{tabular}{lc}
\hline
\textbf{Parameter} & \textbf{Value} \\
\hline
Conductor type & ACSR Drake (795 kcmil) \\
Outer diameter ($D$) & 0.02814 m \\
$R(25^\circ\text{C})$ & $7.283\times10^{-5}$ $\Omega$/m \\
$R(75^\circ\text{C})$ & $8.688\times10^{-5}$ $\Omega$/m \\
$R(200^\circ\text{C})$ & $1.220\times10^{-4}$ $\Omega$/m \\
Max. temp. ($T_{\max}$) & 100 $^\circ$C \\
Emissivity ($\varepsilon$) & 0.8 \\
Solar absorptivity ($\alpha$) & 0.8 \\
Azimuth & 90$^\circ$ (East--West) \\
Altitude & 0 m \\
Latitude ($\phi$) & 30$^\circ$ North \\
Day of year & 161 (June 10) \\
\hline
\end{tabular}
\end{table}

\subsection{Environmental Conditions}

The environmental parameters considered in this study are ambient air temperature, wind speed, and atmospheric clearness. The environmental conditions and study ranges are summarized in Table~\ref{tab:env_conditions}.

Furthermore the IEEE-738-2023 calculates the forced convective heat loss using empirical relations developed by McAdams \cite{mcadams1954}. The two McAdams equations are \eqref{eq:convection1} and \eqref{eq:convection2}.
\begin{equation}
q_{c1} = K_{\text{angle}} \times [1.01 + 1.35 \times N_{Re}^{0.52}] \times k_f \times (T_s - T_a)
\label{eq:convection1}
\end{equation}
\begin{equation}
q_{c2} = K_{\text{angle}} \times 0.754 \times N_{Re}^{0.6} \times k_f \times (T_s - T_a)
\label{eq:convection2}
\end{equation}

Where $N_{Re}$ is the Reynolds number, $K_{\text{angle}}$ is the wind direction factor, $k_f$ is the thermal conductivity of air, $T_s$ is the conductor surface temperature, and $T_a$ is the ambient temperature. The higher value from these two equations is used for conservative DLR calculation. The validated range for these equations are:
\begin{equation}
15~^\circ\text{C} \leq T_a \leq 260~^\circ\text{C}
\end{equation}
\begin{equation}
21~^\circ\text{C} \leq T_s \leq 1004~^\circ\text{C}
\end{equation}
Additional caution is recommended when operating outside these validated ranges to ensure conservative line ratings. Hence ambient temperature is varied from 15~$^\circ$C to 50~$^\circ$C in increments of 5~$^\circ$C. This range aligns with the validated limits of equations \eqref{eq:convection1} and \eqref{eq:convection2}. The lower bound of 15~$^\circ$C corresponds to the minimum validated ambient temperature, while the upper bound of 50~$^\circ$C represents high-temperature conditions in typical Indian climatic scenarios. 

Wind speed is varied from 0~m/s to 15.25~m/s in increments of 0.61~m/s, covering a range representative of normal to moderately high wind speeds encountered in non-cyclonic grid operating conditions. Wind direction is fixed perpendicular to the conductor for all cases, corresponding to an absolute wind direction of 0$^\circ$ incident on a conductor azimuth of 90$^\circ$ (i.e., relative wind direction angle of 90$^\circ$). The perpendicular wind direction ensures maximum convective cooling thereby yielding maximum DLR value.

Atmospheric clearness is represented using two discrete cases corresponding to clear and non-clear sky conditions. Daytime and nighttime conditions are modeled using simulation times of 11:00~IST and 23:00~IST respectively. This allows the effect of solar heat gain to be consistently captured.

\begin{table}[ht]
\caption{Environmental Conditions and Study Ranges}
\label{tab:env_conditions}
\centering
\small
\begin{tabular}{lc}
\hline
\textbf{Parameter} & \textbf{Value / Range} \\
\hline
Ambient temp. ($T_a$) & 15--50 $^\circ$C (step: 5 $^\circ$C) \\
Wind speed ($V_w$) & 0--15.25 m/s (step: 0.61 m/s) \\
Wind direction & 0$^\circ$ (perpendicular) \\
Atm. clearness ($c$) & $\{0, 1\}$ \\
Solar heating - daytime (IST) & 11:00 \\
Solar heating - nighttime (IST) & 23:00 \\
\hline
\end{tabular}
\end{table}

\section{Results and Analysis}

This section presents the results of the study conducted to evaluate the influence of ambient temperature, wind speed, atmospheric clearness, and solar conditions on DLR. For each combination of temperature and wind speed, steady-state DLR is computed, and the analysis is repeated for different atmospheric clearness and day/night conditions. Each operating point corresponds to an independent DLR calculation obtained directly from the heat balance equation.

The resulting dataset is post-processed to examine DLR variation with respect to individual parameters, and to compute sensitivity metrics and Pearson correlation coefficients. Linear regression models with wind--temperature interaction terms are fitted independently for each combination of atmospheric clearness and solar condition to quantify the individual and combined effects of both environmental variables on DLR. Finally, the results are organized into four parts: aggregated DLR variation, sensitivity analysis, correlation analysis, and regression analysis.

\subsection{DLR Variation with Ambient Temperature and Wind Speed}

\begin{figure}[h]
\centering
\subfloat[DLR variation with ambient temperature]{\includegraphics[width=0.45\textwidth]{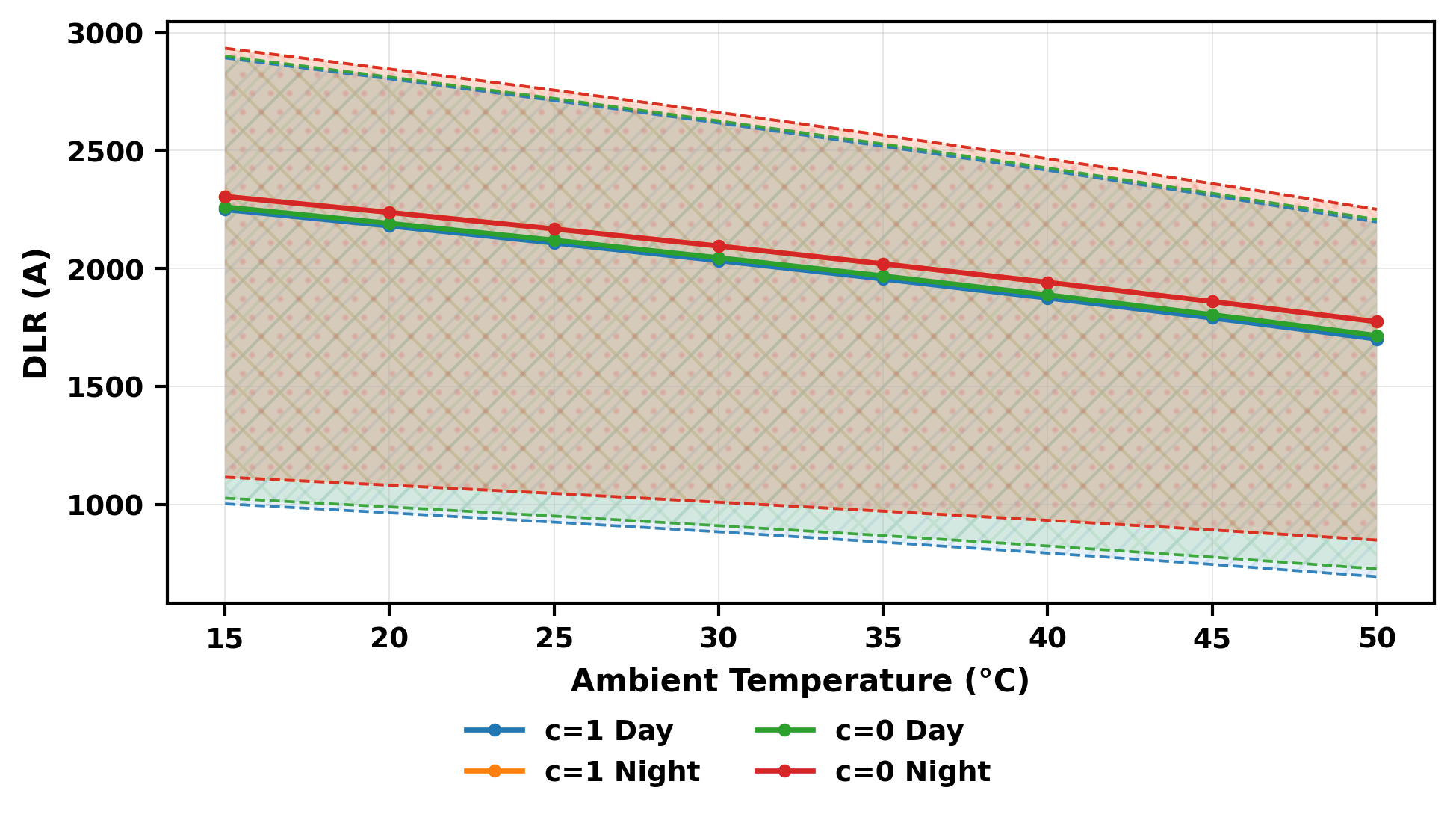}\label{fig:dlr_temp}}
\\
\subfloat[DLR variation with wind speed]{\includegraphics[width=0.45\textwidth]{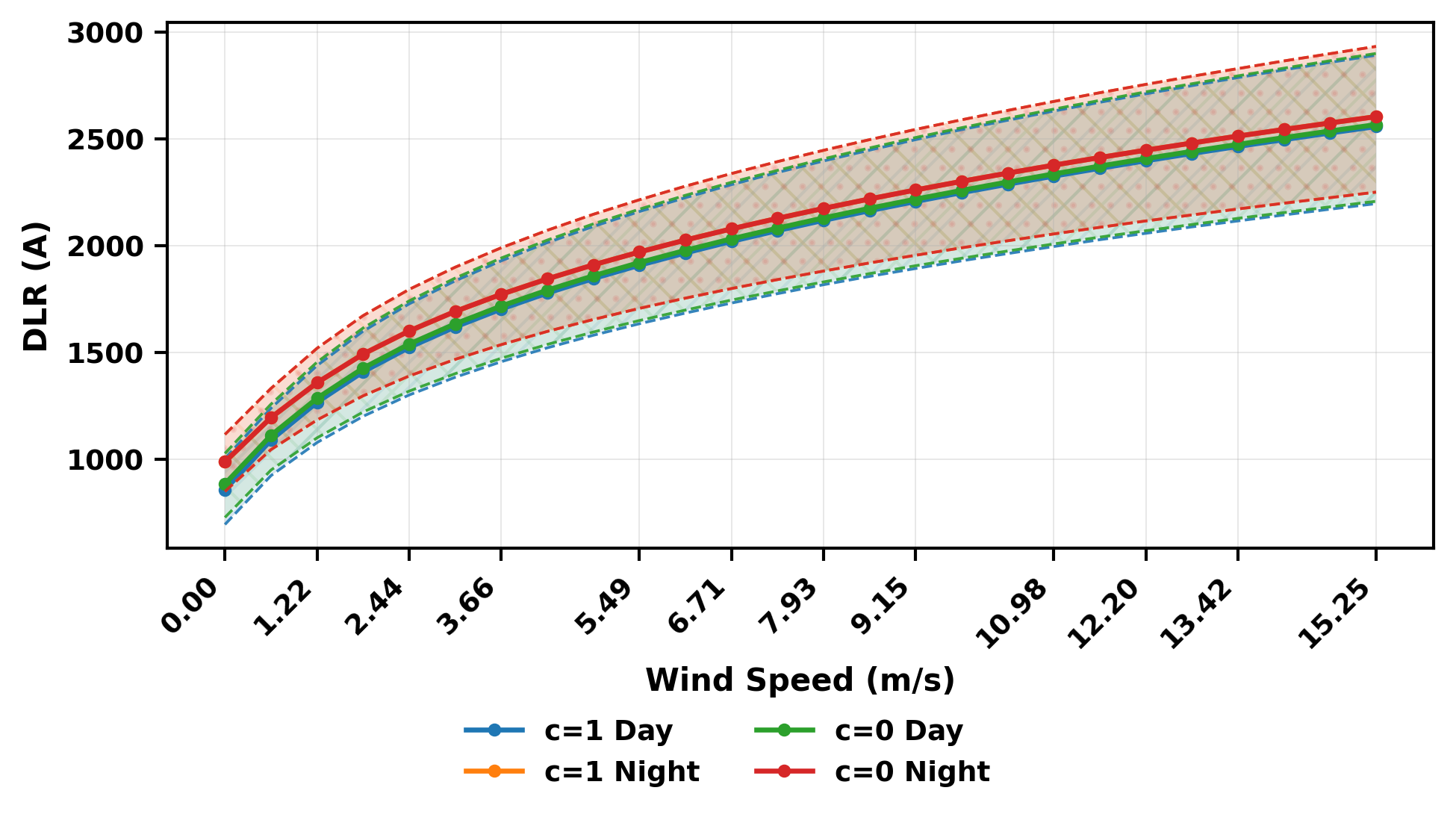}\label{fig:dlr_wind}}
\caption{DLR calculation results under all four atmospheric and solar condition combinations.}
\label{fig:dlr_results}
\end{figure}

Let the steady-state DLR computed using \eqref{eq:dlr} be denoted as $\mathrm{DLR}(T_a, V_w, c, s)$, where $s \in \{\text{day}, \text{night}\}$ represents the solar condition, and all other variables are defined in Table~\ref{tab:env_conditions}. For a fixed ambient temperature $T_a$, the mean DLR aggregated over the wind speed set $\mathcal{V}$ is defined as
\begin{equation}
\overline{\mathrm{DLR}}(T_a, c, s)
=
\frac{1}{|\mathcal{V}|}
\sum_{V_w \in \mathcal{V}}
\mathrm{DLR}(T_a, V_w, c, s),
\end{equation}
with the corresponding minima and maxima given by
\begin{equation}
\mathrm{DLR}_{\min}(T_a, c, s)
=
\min_{V_w \in \mathcal{V}}
\mathrm{DLR}(T_a, V_w, c, s),
\end{equation}
\begin{equation}
\mathrm{DLR}_{\max}(T_a, c, s)
=
\max_{V_w \in \mathcal{V}}
\mathrm{DLR}(T_a, V_w, c, s).
\end{equation}

Fig.~\ref{fig:dlr_temp} shows the mean, minimum, and maximum DLR versus ambient temperature for all four scenarios. At each temperature, the mean curve represents the average over all 26 wind speeds, while the shaded envelope (bounded by min and max) captures the full range of wind-induced variation.

The mean DLR decreases steadily with increasing temperature from 15~$^\circ$C to 50~$^\circ$C for all scenario combinations. The wind-related spread between maximum and minimum DLR values also reduces monotonically as temperature rises. The largest reduction occurs under Nighttime conditions, where the $c = 0$ and $c = 1$ curves overlap completely, with the envelope narrowing from 1819~A at 15~$^\circ$C to 1403~A at 50~$^\circ$C, corresponding to a reduction of 416~A. During Daytime operation, a similar trend is observed, with the envelope decreasing from 1875~A to 1482~A for $c = 0$ and from 1891~A to 1504~A for $c = 1$, indicating a slightly larger reduction for $c = 0$. At all temperatures, the Nighttime mean DLR remains higher than the Daytime mean DLR for both clearness values. Furthermore, the Daytime mean DLR for $c = 0$ remains consistently higher than for $c = 1$, with the difference ranging from 11.423~A to 15.154~A.

Similarly, for a fixed wind speed $V_w$, mean DLR aggregated over the temperature set $\mathcal{T}$ are computed as
\begin{equation}
\overline{\mathrm{DLR}}(V_w, c, s)
=
\frac{1}{|\mathcal{T}|}
\sum_{T_a \in \mathcal{T}}
\mathrm{DLR}(T_a, V_w, c, s),
\end{equation}
with the corresponding minima and maxima given by
\begin{equation}
{\mathrm{DLR}_{\min}}(V_w, c, s)
=
\min_{T_a \in \mathcal{T}}
\mathrm{DLR}(T_a, V_w, c, s),
\end{equation}
\begin{equation}
{\mathrm{DLR}_{\max}}(V_w, c, s)
=
\max_{T_a \in \mathcal{T}}
\mathrm{DLR}(T_a, V_w, c, s),
\end{equation}

Similarly, Fig.~\ref{fig:dlr_wind} shows the mean, minimum, and maximum DLR versus wind speed for all four scenarios. At each wind speed, the mean curve represents the average over all 8 temperature values, while the shaded envelope captures the full range of temperature-induced variation.

The mean DLR increases monotonically with wind speed from 0 to 15.25~m/s for all scenarios, and the temperature-related spread expands progressively as wind speed rises. The largest expansion occurs under Nighttime conditions, where the $c = 0$ and $c = 1$ results overlap completely, with the envelope widening from 267~A at zero wind speed to 683~A at 15.25~m/s, corresponding to an increase of 416~A. During Daytime operation, a similar monotonic increase is observed, with the envelope expanding from 300~A to 693~A for $c = 0$ and from 309~A to 696~A for $c = 1$, indicating slightly larger variability for $c = 0$. Across all wind speeds, the Nighttime mean DLR remains higher than the Daytime mean DLR for both clearness values. The Daytime mean curves remain closely spaced; however, the $c = 0$ Daytime mean DLR is consistently higher than the $c = 1$ Daytime mean DLR, with the difference varying between 9.250~A and 27.875~A. 

Since solar heat gain will be $0$ during the night, both Nighttime cases (i.e., at $c = 0$ \& $c = 1$) overlap. Thus, only three visually distinct curves appear in Fig.~\ref{fig:dlr_results}.

\subsection{Sensitivity Analysis}

Temperature sensitivity $S_T$ and wind sensitivity $S_V$ are respectively defined as:
\begin{equation}
S_T =
\frac{\Delta \mathrm{DLR}}{\Delta T_a},
\end{equation}
\begin{equation}
S_V =
\frac{\Delta \mathrm{DLR}}{\Delta V_w}.
\end{equation}

\begin{figure}[h]
\centering
\subfloat[Temperature sensitivity]{\includegraphics[width=0.45\textwidth]{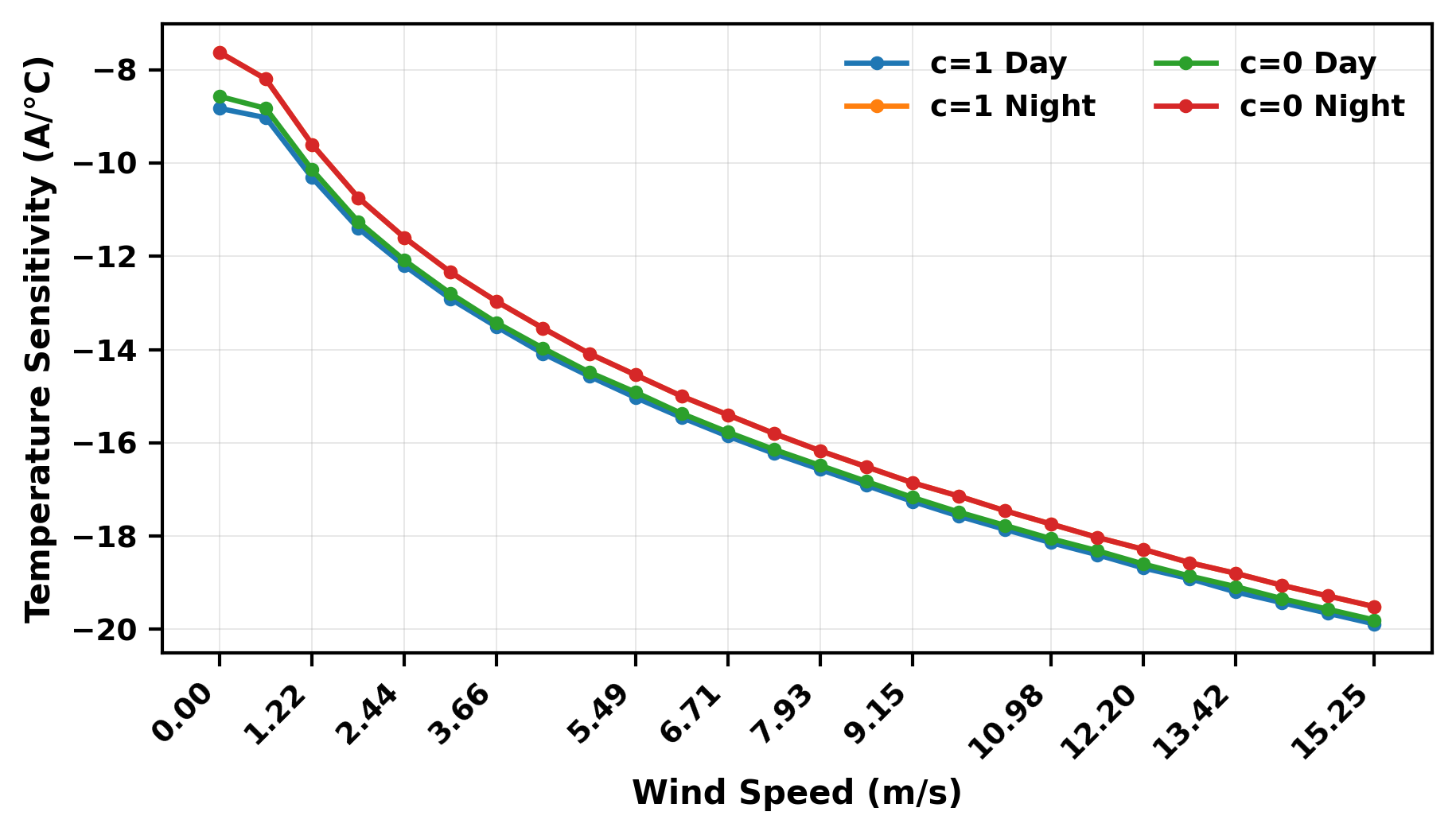}\label{fig:temp_sensitivity}}
\\
\subfloat[Wind sensitivity]{\includegraphics[width=0.45\textwidth]{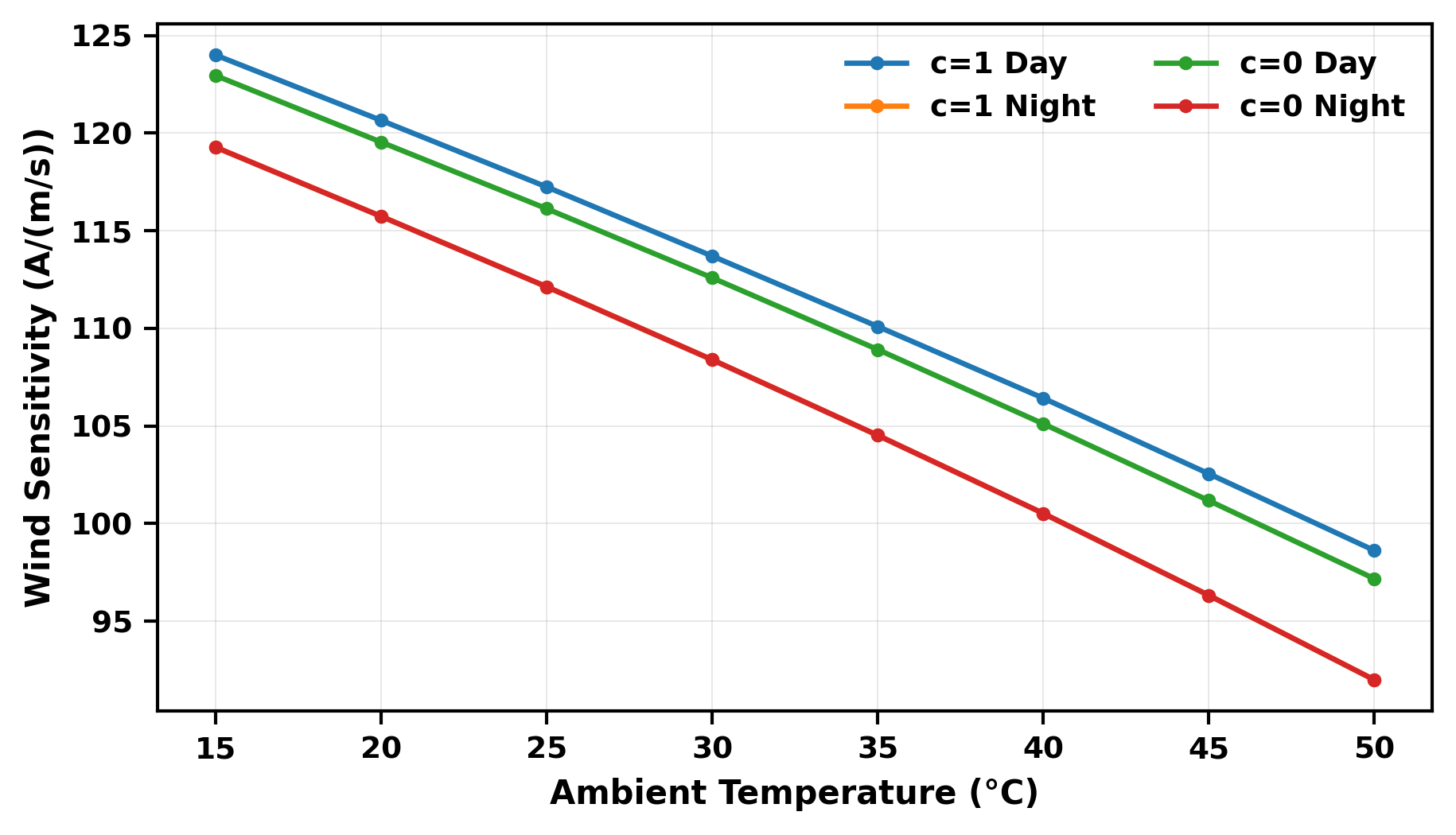}\label{fig:wind_sensitivity}}
\caption{Sensitivity analysis results under all four atmospheric and solar condition combinations.}
\label{fig:sensitivity_results}
\end{figure}

Fig.~\ref{fig:temp_sensitivity} shows temperature sensitivity versus wind speed. Each point represents the average rate of change computed via finite differences across 7 temperature intervals (15--50~$^\circ$C) at fixed wind speed, clearness, and solar condition.

The temperature sensitivity becomes progressively more negative as wind speed increases from 0 to 15.25~m/s for all scenario combinations. At every wind speed, the sensitivity under Nighttime conditions is less negative than under Daytime conditions for both $c = 0$ and $c = 1$. For both Daytime and Nighttime cases, the $c = 0$ sensitivity remains less negative than the $c = 1$ sensitivity across the entire wind speed range. Throughout the wind speed range, the Daytime curves for $c = 0$ and $c = 1$ remain very close, with the difference varying from 0.0571 to 0.2571~A/$^\circ$C, and the $c = 0$ Daytime curve consistently exhibiting slightly less negative values.

Fig.~\ref{fig:wind_sensitivity} shows wind sensitivity versus temperature. Each point represents the average rate of change computed via finite differences across 25 wind speed intervals (0--15.25~m/s) at fixed temperature, clearness, and solar condition.

The wind sensitivity decreases consistently as temperature increases from 15~$^\circ$C to 50~$^\circ$C for all scenario combinations. At every temperature, the wind sensitivity under Daytime conditions is higher than under Nighttime conditions for both $c = 0$ and $c = 1$. During Daytime operation, the $c = 1$ sensitivity remains higher than the $c = 0$ sensitivity throughout the temperature range, with the difference varying from $-1.4426$ to $-1.0492$~A/(m/s), indicating that $c = 1$ exceeds $c = 0$ by approximately 1.05 to 1.44~A/(m/s). Across the full temperature range, the highest wind sensitivity values occur for the $c = 1$ Daytime case, decreasing from 124.0~A/(m/s) at 15~$^\circ$C to 98.623~A/(m/s) at 50~$^\circ$C, while the lowest values occur under Nighttime conditions, decreasing from 119.279~A/(m/s) at 15~$^\circ$C to 92.0~A/(m/s) at 50~$^\circ$C.

The two Nighttime curves corresponding to $c = 0$ and $c = 1$ overlap exactly at all wind speeds, resulting in fewer visually distinct curves in Fig.~\ref{fig:sensitivity_results}

\subsection{Correlation Analysis}

\begin{figure}[h]
\centering
\subfloat[DLR-wind speed correlation at different temperatures]{\includegraphics[width=0.45\textwidth]{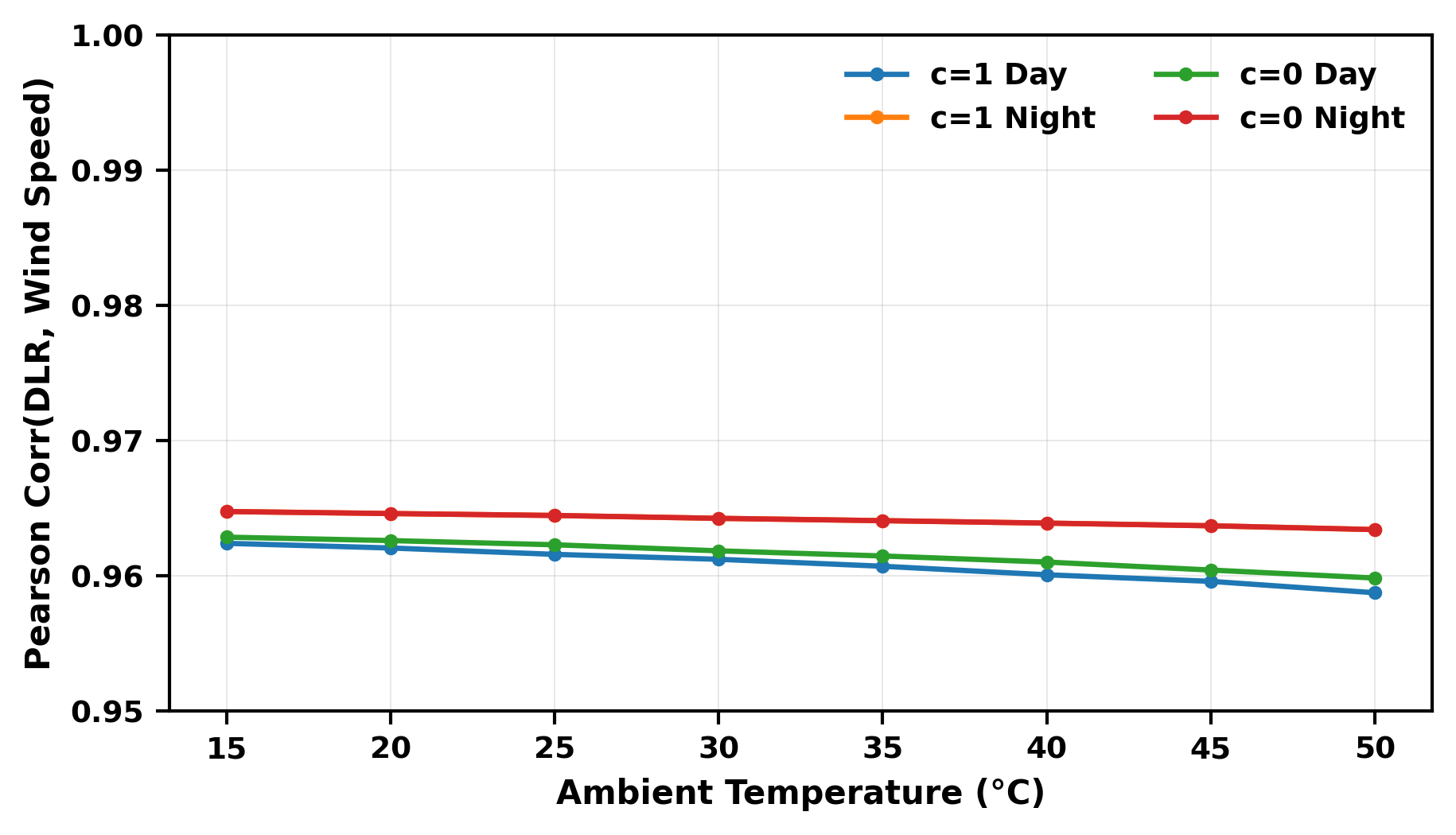}\label{fig:corr_dlr_wind}}
\\
\subfloat[DLR-temperature correlation at different wind speeds]{\includegraphics[width=0.45\textwidth]{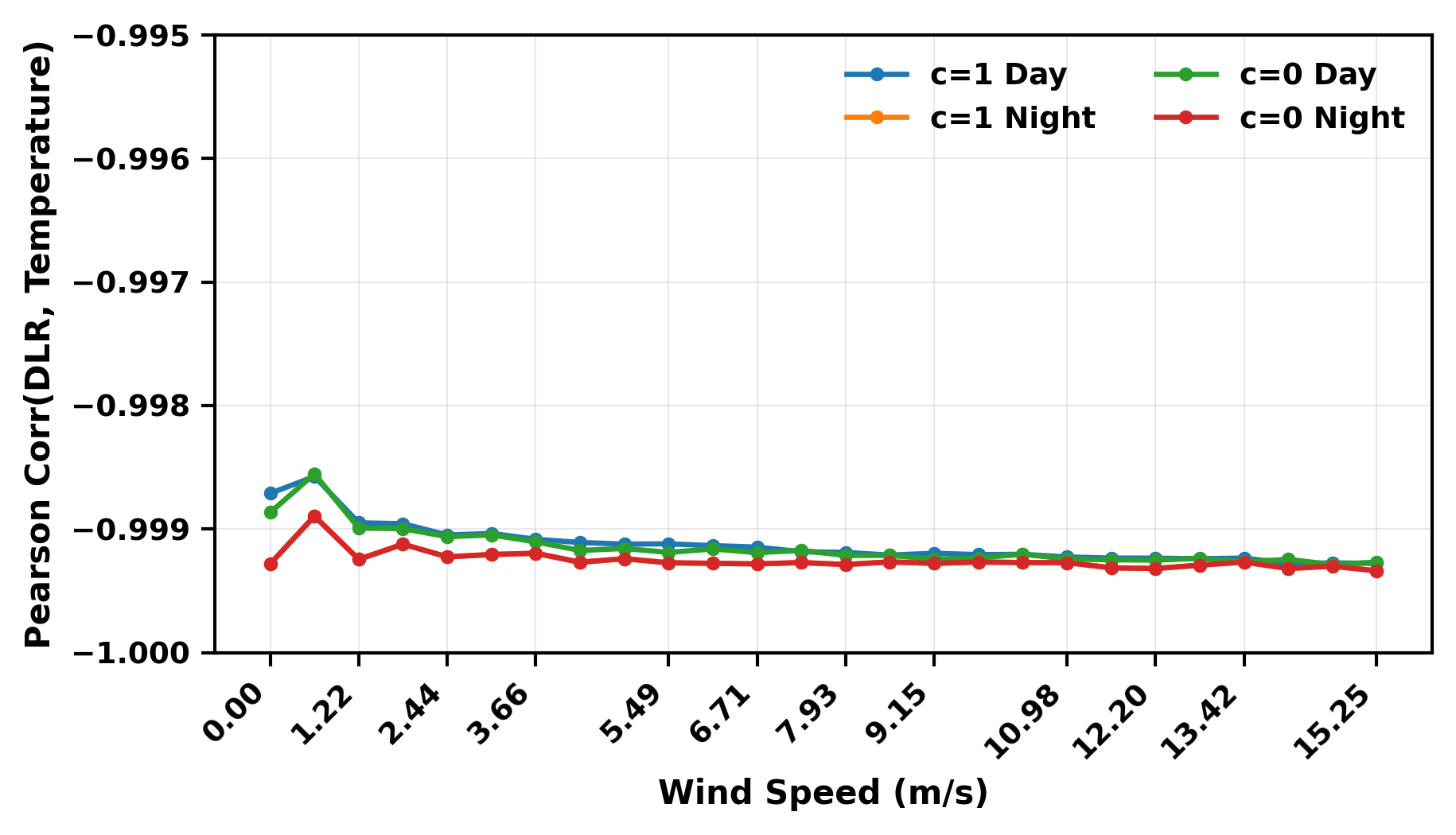}\label{fig:corr_dlr_temp}}
\caption{Pearson correlation coefficients under all four atmospheric and solar condition combinations.}
\label{fig:corr_dlr_results}
\end{figure}

Pearson correlation analysis is performed to quantify the strength of association between DLR and the environmental variables. The results are presented in Fig.~\ref{fig:corr_dlr_results}.

Fig.~\ref{fig:corr_dlr_wind} shows the Pearson correlation between DLR and wind speed at each ambient temperature. Each point is computed using 26 DLR values corresponding to 26 wind speeds at fixed temperature, atmospheric clearness, and solar condition. The correlation remains strongly positive for all scenario combinations over the entire temperature range from 15~$^\circ$C to 50~$^\circ$C. For each scenario, the correlation decreases steadily as ambient temperature increases, indicating a gradual reduction in the strength of the wind-DLR association at higher temperatures. At every temperature, the correlation under Nighttime conditions is higher than under Daytime conditions for both $c = 0$ and $c = 1$. During Daytime operation, the $c = 0$ correlation remains consistently higher than the $c = 1$ correlation across all temperatures, with the difference varying between 0.000456 and 0.001088.

Fig.~\ref{fig:corr_dlr_temp} shows the Pearson correlation between DLR and ambient temperature at each wind speed. Each point is computed using 8 DLR values corresponding to 8 temperatures at fixed wind speed, atmospheric clearness, and solar condition. The correlation remains strongly negative and very close to $-1$ for all scenario combinations across the full wind speed range. Although small fluctuations are observed with changes in wind speed, no clear monotonic trend is evident for any scenario. At every wind speed, the correlation under Nighttime conditions is more negative than under Daytime conditions for both $c = 0$ and $c = 1$. During Daytime operation, the $c = 0$ and $c = 1$ curves remain extremely close throughout, with the maximum difference between them on the order of $1.55 \times 10^{-4}$.

Similar to earlier subsections, the Nighttime curves for $c = 0$ and $c = 1$ overlap exactly, indicating identical correlation values in Fig.~\ref{fig:corr_dlr_results}.

\subsection{Regression Analysis of Combined Effects}

To quantify the individual and combined influence of ambient temperature and wind speed on DLR, linear regression models with wind--temperature interaction terms are fitted independently for each scenario. The regression model is expressed in \eqref{eq:regression}, wherein $\bar{T}_a = 32.5~^\circ$C and $\bar{V}_w = 7.625$~m/s are the mean $T_a$ \& $V_w$ over the dataset. The estimated coefficients and goodness-of-fit metrics are reported in Table~\ref{tab:regression_results}.
\begin{equation}
\begin{aligned}
\mathrm{DLR} ={}& \beta_0 
+ \beta_{T_a}(T_a - \bar{T}_a) 
+ \beta_{V_w}(V_w - \bar{V}_w) \\
&+ \beta_{T_a V_w}(T_a - \bar{T}_a)(V_w - \bar{V}_w)
\label{eq:regression}
\end{aligned}
\end{equation}

\begin{table}[ht]
\caption{Regression Results of Combined Effects on DLR}
\label{tab:regression_results}
\centering
\small
\begin{tabular}{cccccccc}
\hline
$c$ & $s$ & $n$ & $\beta_0$ & $\beta_{T_a}$ & $\beta_{V_w}$ & $\beta_{T_a V_w}$ & $R^2$ \\
\hline
0 & D & 208 & 1999.32 & -15.54 & 95.29 & -0.695 & 0.935 \\
0 & N & 208 & 2050.19 & -15.15 & 92.43 & -0.718 & 0.939 \\
1 & D & 208 & 1986.30 & -15.65 & 96.07 & -0.689 & 0.933 \\
1 & N & 208 & 2050.19 & -15.15 & 92.43 & -0.718 & 0.939 \\
\hline
\end{tabular}
\end{table}

The intercept $\beta_0$ represents predicted DLR at mean environmental conditions (32.5~$^\circ$C, 7.625~m/s), ranging from 1986.30~A ($c = 1$, daytime) to 2050.19~A (both nighttime cases). The Nighttime coefficients for $c = 0$ and $c = 1$ are identical, consistent with the absence of solar heating in both cases. The temperature coefficient $\beta_{T_a}$ ranges from $-15.65$ to $-15.15$~A/$^\circ$C, confirming that at mean wind speed, each 1~$^\circ$C temperature increase reduces DLR by approximately 15--16~A. The wind coefficient $\beta_{V_w}$ ranges from 92.43 to 96.07~A/(m/s), indicating that at mean temperature, each 1~m/s wind increase raises DLR by approximately 92--96~A. The interaction coefficient $\beta_{T_a V_w}$ is consistently negative ($-0.689$ to $-0.718$), capturing the coupled effect of both parameters. This negative interaction indicates that wind cooling effectiveness diminishes at higher temperatures, while temperature sensitivity amplifies at higher wind speeds. The coefficient of determination ($R^2$) ranges from 0.933 to 0.939, confirming that the linear model with interaction explains over 93\% of observed DLR variability across all scenarios.

\section{Conclusion}

This paper presented a systematic multi-scenario sensitivity analysis of dynamic line rating for a 795~kcmil ACSR Drake conductor using the IEEE-738-2023 standard. DLR was evaluated over a structured set of operating points that covers practical ranges of ambient temperature and wind speed under four environmental scenarios based on atmospheric clearness and solar condition combinations.

The results across all four scenarios show that ambient temperature and wind speed jointly govern DLR behavior in a coupled manner. DLR decreases consistently with increasing temperature and increases nonlinearly with wind speed. Temperature sensitivity strengthens at higher wind speeds, while wind sensitivity decreases at higher ambient temperatures. Correlation analysis confirmed a strong negative linear association between DLR and temperature, along with a strong positive association with wind speed.

Linear regression models explained over 93\% of observed DLR variability across all atmospheric and solar scenarios. The estimated temperature and wind coefficients quantify how each parameter directly contributes to DLR, while the consistently negative interaction term captures their coupled behavior—specifically, wind cooling becomes less effective at higher ambient temperatures, and temperature sensitivity grows stronger under higher wind conditions.

Overall, this study offers a quantitative description of DLR behavior across realistic ranges of wind speed and ambient temperature under various atmospheric and solar conditions. The variation, sensitivity, and regression analyses that are provided give operators a structured understanding of how DLR variability is influenced by uncertainties in important weather parameters. Therefore, these relationships can guide power system operators to integrate DLR into grid operations and planning more intelligently in situations where environmental fluctuations may significantly affect thermal ratings.

\bibliographystyle{IEEEtran}
\bibliography{references}

\end{document}